\documentclass[preprint, review, 12pt, 3p]{elsarticle}

\usepackage{amsmath}
\usepackage{dirtytalk}
\usepackage[hidelinks]{hyperref}
\usepackage{graphicx}
\usepackage{subfig}
\usepackage{multirow}
\usepackage{lipsum}
\usepackage{natbib}
\usepackage{longtable}
\usepackage{caption}
\usepackage{adjustbox}
\usepackage{tabularx}
\usepackage{floatrow} 
\usepackage[vlines]{tabularht}
\usepackage{pbox}

\usepackage{float}
\floatstyle{plaintop}
\restylefloat{table}

\usepackage[linesnumbered,ruled,vlined]{algorithm2e}
\SetKwInput{KwInput}{Input}                
\SetKwInput{KwOutput}{Output}              

\usepackage[table,xcdraw]{xcolor}

\newsavebox\MBox

\biboptions{sort&compress, square}

\journal{Artificial intelligence in medicine}

\begin{document}

\begin{frontmatter}

\title{COVID-19 identification from volumetric chest CT scans using a progressively resized 3D-CNN incorporating segmentation, augmentation, and class-rebalancing}

\author[label1]{Md. Kamrul Hasan\corref{cor1}\fnref{label3}}
\address[label1]{Department of Electrical and Electronic Engineering, Khulna University of Engineering \& Technology, Khulna-9203, Bangladesh}
\address[label6]{Department of Computer Science and Engineering, Khulna University of Engineering \& Technology, Khulna-9203, Bangladesh}
\address[label51]{University Health Network, Toronto, ON, Canada}
\address[label52]{Institute of Biomedical Engineering, University of Toronto, Toronto, ON, Canada}
\cortext[cor1]{I am corresponding author}
\fntext[label4]{Department of EEE, KUET, Khulna-9203, Bangladesh.}
\ead{m.k.hasan@eee.kuet.ac.bd}

\author[label1]{Md. Tasnim Jawad}
\ead{jawad1703006@stud.kuet.ac.bd}

\author[label6]{Kazi Nasim Imtiaz Hasan}
\ead{hasan1607097@stud.kuet.ac.bd}

\author[label6]{Sajal Basak Partha}
\ead{partha1607101@stud.kuet.ac.bd}

\author[label6]{Md. Masum Al Masba}
\ead{masba@cse.kuet.ac.bd}

\author[label51,label52]{Shumit Saha}
\ead{shumit.saha@mail.utoronto.ca}

\begin{abstract}
The novel COVID-19 is a global pandemic disease overgrowing worldwide. 
Computer-aided screening tools with greater sensitivity is imperative for disease diagnosis and prognosis as early as possible. It also can be a helpful tool in triage for testing and clinical supervision of COVID-19 patients.
However, designing such an automated tool from non-invasive radiographic images is challenging as many manually annotated datasets are not publicly available yet, which is the essential core requirement of supervised learning schemes. 
This article proposes a 3D Convolutional Neural Network (CNN)-based classification approach considering both the inter- and intra-slice spatial voxel information. 
The proposed system is trained in an end-to-end manner on the 3D patches from the whole volumetric CT images to enlarge the number of training samples, performing the ablation studies on patch size determination. 
We integrate progressive resizing, segmentation, augmentations, and class-rebalancing to our 3D network. The segmentation is a critical prerequisite step for COVID-19  diagnosis enabling the classifier to learn prominent lung features while excluding the outer lung regions of the CT scans. 
We evaluate all the extensive experiments on a publicly available dataset, named MosMed, having binary- and multi-class chest CT image partitions. 
Our experimental results are very encouraging, yielding areas under the ROC curve of $0.914\pm 0.049$ and $0.893 \pm 0.035$ for the binary- and multi-class tasks, respectively, applying 5-fold cross-validations.
Our method's promising results delegate it as a favorable aiding tool for clinical practitioners and radiologists to assess COVID-19.\\ 

\end{abstract}


\end{frontmatter}

\section{Introduction}
\label{introduction}
Pneumonia of unknown cause discovered in Wuhan, China, was published to the World Health Organization (WHO) office in China on 31st December 2019. It was consequently assigned to severe acute respiratory syndrome coronavirus 2 (SARS-CoV-2) because of having similar genetic properties to the SARS outbreak of 2003. On 11th February 2020, WHO termed that new disease as COVID-19 (Coronavirus disease), which displays an upper respiratory tract and lung infection \citep{naming_covid}. The clinical characteristics of critical COVID-$19$ pandemic are bronchopneumonia that affects cough, fever, dyspnea, and detailed respiratory anxiety ailment \citep{wang2020clinical, chen2020epidemiological, li2020early}. 
According to the WHO reports, COVID-19's general indications are equivalent to that of ordinary flu, including fever, tiredness, dry cough, shortness of breath, aches, pains, and sore throat \citep{jain2020deep}. 
Those shared signs turn it challenging to recognize the virus at an ancient step. The aforementioned is a virus, which works on bacterial or fungal infections \citep{wu2020outbreak, jain2020deep} with no possibility that antibiotics can restrict it. Besides, people suffering from medical complications, like diabetes, chronic respiratory and cardiovascular diseases, are prone to undergo COVID-19. 
An explanatory statement of the Imperial College advised that the affection rate will be more than $90.0\,\%$ of the world’s people, killing 40.6 million people if no reduction actions are grasped to combat the virus \citep{walker2020impact}.

Advanced presumed discovery of COVID-19 is also a challenge for public health security and control of pandemic. The COVID-19 detection failure increases the mortality rate exponentially. 
The incubation period, which is a time between catching the virus and causing to have indications of the illness, is $1\sim14$ days, making it remarkably challenging to identify COVID-19 infection at a preliminary stage of an individual's symptoms \citep{jain2020deep}.
The clinical screening test for the COVID-$19$ is Reverse Transcription Polymerase Chain Reaction (RT-PCR), practicing respiratory exemplars. 
However, it is a manual, complicated, tiresome, and time-consuming fashion with an estimated true-positive rate of $63.0\%$ \citep{wang2020detection}.
There is also a significant lack of RT-PCR kit inventory, leading to a delay in preventing and curing coronavirus disease \citep{yang2020point}.
Furthermore, the RT-PCR kit is estimated to cost around $120\sim 130$ USD. It also requires a specially designed biosafety laboratory to house the PCR unit, each of which can cost $15,000\sim 90,000$ USD  \citep{aj2020corona}. 
Nevertheless, the utilization of a costly screening device with a delayed test results makes it more challenging to restrict the disease's spread.
Inadequate availability of screening workstations and measurement kits constitute an enormous hardship to identify COVID-19 in this pandemic circumstance. In such a situation, speedy and trustworthy presumed COVID-19 cases are an enormous difficulty for related personals.

However, it is observed that most of the COVID-$19$ incidents have typical properties on radiographic CT and X-ray images, including bilateral, multi-focal, ground-glass opacities with a peripheral or posterior distribution, chiefly in the lower lobes and early- and late-stage pulmonary concentration \citep{huang2020clinical, corman2020detection,xu2020deep,singh2020classification}. 
Those features can be utilized to build a sensitive Computer-aided Diagnosis (CAD) tool to identify COVID-$19$ pneumonia, which is deemed an automated screening tool \citep{lee2020covid}. 
Currently, deep Convolutional Neural Networks (CNNs) allow for building an end-to-end model without requiring manual and time-consuming feature extraction and engineering \citep{lecun2015deep, krizhevsky2012imagenet}, demonstrating tremendous success in many domains of medical imaging, such as arrhythmia detection \citep{yildirim2018arrhythmia,hannun2019cardiologist,acharya2017deep}, skin lesion segmentation and classification \citep{hasan2020dsnet, esteva2017dermatologist, codella2017deep, dutta2020skin}, breast cancer detection \citep{celik2020automated, cruz2014automatic, hasan2019automatic}, brain disease segmentation and classification \citep{talo2019convolutional,tushar2019brain}, pneumonia detection from chest X-ray images \citep{rajpurkar2017chexnet}, fundus image segmentation \citep{tan2017automated,hasan2020drnet}, and lung segmentation \citep{gaal2020attention}.
Most recently, various deep CNN-based methods have been published for identifying COVID-19 from X-rays and CT images, summarizing and bestowing in Table~\ref{tab:previous_methods_with_results}. 
\begin{table*}[!ht]
\caption{Numerous published articles for the COVID-19 identification with their respective utilized datasets and performances exhibiting different metrics such as mSn, mSp, and mF1 respectively for mean sensitivity, specificity, and F1-score. The mixed datasets indicate that data have come from different open-sources. }
\centering
\tiny
\begin{tabular}{p{11.8cm}ll}
\hline
\rowcolor[HTML]{C0C0C0} 
Different methods & Datasets & Results \\ \hline

\multirow{2}{11.8cm}{A pre-trained 2D MobileNet-v2 \citep{sandler2018mobilenetv2} architecture on ImageNet \citep{deng2009imagenet} was used to extract massive high-dimensional features to classify six different diseases using the fully-connected layers \citep{apostolopoulos2020extracting}  }     &                     \multirow{2}*{Mixed}            &         mSn: $0.974$                                   \\
 &         &   mSp: $0.994$                                     \\ \hline

\multirow{2}{11.8cm}{DeTraC \citep{abbas2020detrac, abbas2020classification}, where the network was trained first using a gradient descent optimization \citep{ruder2016overview}, and then, the class-composition layer of DeTraC was used to refine the final detection results \citep{abbas2020classification} }     &                     \multirow{2}*{Mixed}            &         mSn: $0.979$                                   \\
 &         &   mSp: $0.919$                                     \\ \hline

\multirow{2}{11.8cm}{A multi-objective differential evolution–based CNN method fine-tuning iteratively using mutation, crossover, and selection operations to discover the best possible results \citep{singh2020classification}}     &                     \multirow{2}*{Mixed}            &         mSn: $0.907$                                   \\
 &         &   mSp: $0.906$                                     \\ \hline

\multirow{2}{11.8cm}{An ensemble of VGG-16 \citep{simonyan2014very}, Inception \citep{szegedy2015going}, Xception \citep{chollet2017xception}, Inception-ResNet \citep{szegedy2016inception}, MobileNet \citep{howard2017mobilenets}, DenseNet \citep{huang2017densely}, and NasNet \citep{pham2018efficient} optimizing the hyperparameters using a greedy search algorithm \citep{bergstra2012random,rajaraman2020iteratively} }     &                     \multirow{2}*{Mixed}            &         mSn: $0.990$                                   \\
 &         &   mSp: $0.990$                                     \\ \hline

\multirow{2}{11.8cm}{Support vector machine \citep{wu2012combining, hasan2019automatic}-based method to classify the in-depth features from the pre-trained MobileNet and SqueezeNet \citep{iandola2016squeezenet} from the restructured the data using a fuzzy color technique \citep{tougaccar2020covid}}     &                     \multirow{2}*{Mixed}            &         mSn: $0.983$                                   \\
 &         &   mSp: $0.997$                                     \\ \hline

\multirow{2}{11.8cm}{An ensemble of three lightweight pre-trained SqueezeNet, ShuffleNet \citep{zhang2018shufflenet}, and EfficientNet-B0 \citep{tan2019efficientnet} at various depths and consolidates feature maps in diverse abstraction levels \citep{oksuz2020ensemble}}     &                     \multirow{2}*{Mixed}            &         mSn: $0.978$                                   \\
 &         &   mSp: $0.985$                                     \\ \hline   

\multirow{2}{11.8cm}{A fusing and ranking of in-depth features for classifying using a support vector machine, where the pre-trained CNN models on ImageNet were used to extract the COVID-19 features \citep{ozkaya2020coronavirus}}     &                     \multirow{2}*{Mixed}            &         mSn: $0.989$                                   \\
 &         &   mSp: $0.976$                                     \\ \hline

\multirow{2}{11.8cm}{A DenseNet-201 \citep{huang2017densely}-based transfer learning to extract features using its learned weights on the ImageNet was used to classify the patients as COVID infected or not  \citep{jaiswal2020classification} }     &                     \multirow{2}*{SARS-COV-2 \citep{angelov2020explainable}}
&         mSn: $0.960$                                   \\
 &         &   mSp: $0.960$                                     \\ \hline

\multirow{2}{11.8cm}{A transfer learning-based approach using one of the VGG, ResNet \citep{he2016deep}, Inception, or Xception pre-trained deep learning model on ImageNet as a backbone \citep{ko2020covid}}     &                     \multirow{2}*{\citet{zhao2020covid}}            &         mSn: $0.996$    
\\
 &         &   mSp: $0.100$                                     \\ \hline

 \multirow{2}{11.8cm}{A weakly-supervised learning schema, where the lung region was segmented using a pre-trained UNet \citep{ronneberger2015u}; then, a 3D network was used to predict the probability of COVID-19 infectious \citep{wang2020weakly}}     &                     \multirow{2}*{\citet{wang2020weakly}}  
 
 &         mSn: $0.911$                                   \\

 &         &   mSp: $0.881$                                     \\ \hline

 \multirow{2}{11.8cm}{A multi-scale-multi-encoder ensemble of CNN model aggregating the outputs from two different encoders and their different scales to obtain the final prediction probability \citep{hasan2020cvr}}     &                     \multirow{2}*{Mixed}            &         mSn: $0.997$                                   \\
 &         &   mSp: $0.997$                                     \\ \hline

 \multirow{2}{11.8cm}{Advanced deep network architectures proposing a transfer learning strategy on ImageNet using a custom-sized input tailored for each architecture to achieve the best possible results \citep{alshazly2020explainable}}     &                     \multirow{2}*{Mixed}            &         mSn: $0.996$                                   \\
 &         &   mSp: $0.998$                                     \\ \hline

 \multirow{2}{11.8cm}{A pre-trained CNN-based schema leveraging the strength of multiple texture descriptors and base classifiers at once, where data was re-balanced using resampling algorithms \citep{pereira2020covid}}  &                     \multirow{2}*{Mixed}            &         mSn: $-$                                   \\
 &         &   mF1: $0.889$                                     \\ \hline 
 
 \multirow{2}{11.8cm}{A deep ResNet-based transfer learning technique with a top-2 smooth loss function and a cost-sensitive attribute to handle noisy and imbalanced COVID-19 datasets \citep{pathak2020deep}}     &                     \multirow{2}*{Mixed}            &         mSn: $0.915$                                   \\
 &         &   mSp: $0.948$                                     \\ \hline

 \multirow{2}{11.8cm}{An auxiliary classifier generative adversarial network-based design to generate synthetic images, where the synthetic images produced CNN's enhanced results for the prediction \citep{waheed2020covidgan}}     &                     \multirow{2}*{Mixed}            &         mSn: $0.900$                                   \\

 &         &   mSp: $0.970$                                     \\ \hline

 \multirow{2}{11.8cm}{A framework consisting of a CNN-based feature extractor and k-nearest neighbor \citep{hasan2017prediction,hasan2020diabetes}, support vector machine, and decision tree \citep{hasan2020diabetes}-based classifiers using the Bayesian algorithm \citep{nour2020novel}}     &                     \multirow{2}*{Mixed}            &         mSn: $0.894$                                   \\

 &         &   mSp: $0.998$                                     \\ \hline 
 
 \multirow{2}{11.8cm}{An architecture based on the deep residual neural network using two parallel levels with different kernel sizes for capturing both local and global features of the inputs images \citep{ouchicha2020cvdnet}}     &                     \multirow{2}*{Mixed}            &         mSn: $-$                                   \\

 &         &   mF1: $0.967$                                     \\ \hline 
 
 \multirow{2}{11.8cm}{A classification architecture combining ResNet and Xception to investigate the challenges and limitations of deep CNN and different datasets for building generic COVID-19 classifiers   \citep{hasan2020challenges} }     &                     \multirow{2}*{Mixed}            &         mSn: $0.976$                                   \\

 &         &   mSp: $-$                                     \\ \hline

 \multirow{2}{11.8cm}{An average rank pooling, multiple-way augmentation, and deep feature fusion-based CNN and graph CNN was developed to fuse individual image-level features and relation-aware features \citep{wang2020covid} }     &                     \multirow{2}*{\citet{wang2020covid}}            &         mSn: $0.963$                                   \\

 &         &   mSp: $0.970$                                     \\ \hline

 \multirow{2}{11.8cm}{An end-to-end DarkCovidNet architecture \citep{ozturk2020automated} based on DarkNet \citep{ozturk2020automated} gradually increasing the number of filters, where each convolutional layers were followed by BatchNorm \citep{ioffe2015batch} and LeakyReLU \citep{xu2015empirical}  }     &                     \multirow{2}*{Mixed}            &         mSn: $0.951$                                   \\

 &         &   mSp: $0.953$                                     \\ \hline 
 
 \multirow{2}{11.8cm}{A CoroNet model based on pre-trained Xception architecture on ImageNet for automated detection of COVID-19 infection and trained in end-to-end manners \citep{khan2020coronet} }     &                     \multirow{2}*{Mixed}            &         mSn: $0.993$                                   \\

 &         &   mSp: $0.986$                                     \\ \hline 
 
 \multirow{2}{11.8cm}{Comparative analyses of different pre-trained models considering several important factors such as batch size, learning rate, epoch numbers, and type of optimizers to find the best-suited model  \citep{nayak2020application} }     &                     \multirow{2}*{Mixed}            &         mSn: $0.100$                                   \\
 &         &   mSp: $0.967$  
 \\ \hline

 \multirow{3}{11.8cm}{A comparative analysis of different CNN models, such as VGG, Resnet, Inception, Xception, Inception-ResNet, DenseNet, and NASNet-Large \citep{zoph2018learning}, to decide a proper one for multi-modal image classification minimizing the image quality imbalances in the image samples as  a preprocessing \citep{horry2020covid}}     &                     \multirow{3}*{Mixed}            &         mSn: $0.820$                                   \\

 &         &   mSp: $-$  \\       
&         &   mF1: $0.820$    
 \\ \hline 
 
\multirow{2}{11.8cm}{A pipeline consisting of segmentation and subsequent classification employing both 3D and 2D CNNs, where the promising results for detecting were obtained in the 3D-CNNs than the 2D CNNs   \citep{he2020benchmarking} }     &                     \multirow{2}*{\citet{he2020benchmarking}}            &         mSn: $0.891$                                   \\

 &         &   mSp: $0.911$                                     \\ \hline

\end{tabular}
\label{tab:previous_methods_with_results}
\end{table*}
Though the results obtained in the current articles are promising, they exhibit limited scope for use as a CAD tool, as most of the works, especially on x-ray images, have been based on data coming from different sources for two distinct classes (Covid Vs. Normal) \cite{singh2020classification,apostolopoulos2020covid,wang2020covid,sethy2020detection,hemdan2020covidx,narin2020automatic,ozturk2020automated,khan2020coronet}. 
This brings inherent bias on the algorithms as the model tends to learn the distribution of the data source for binary classification problem \citep{hasan2020challenges}. Therefore, these models perform very low when used in practical settings, where the models have to adapt to data from different domains \citep{hasan2020challenges}. 
Recently, \citet{morozov2020mosmeddata}  launched a public chest volumetric CT scan dataset with 1110 COVID-19 related studies (see details in subsection~\ref{Dataset}). However, the published articles \citep{mahmud2021covtanet,yip2020performance} on this dataset consider only intra-slice spatial voxel information to isolate COVID-19 and regular healthy patients.

This article aims to evaluate the proposed 3D-CNN classifier's performance for identifying COVID-19 utilizing volumetric chest images, where the volumes have come from the same source (details in subsection~\ref{Dataset}). However, the core contributions in this article are enlisted as follows: 

\begin{itemize}
  \item Designing a 3D-CNN-based classification network for volumetric CT images as the 3D networks account for the inter- and intra-slice spatial voxel information while the 2D networks consider only the intra-slice spatial voxel information \citep{jnawali2018deep,singh20203d,huang2017lung,zhou2018performance,he2020benchmarking,yip2020performance}
  
  \item Conducting 3D patch-based classification as it increases the sample numbers in the smaller datasets, where we perform ablation studies to determine a proper patch size
  
  \item Progressively increasing the input patch size of our network up to the original CT size of $R\times C \times S$, where the trained network with the patch size of $(R/2^{n+1})\times (C/2^{n+1}) \times (S/2^{n+1})$ is a pre-trained model of a network with the patch size of $(R/2^{n})\times (C/2^{n}) \times (S/2^{n})$  
  
  \item Developing an unsupervised lung segmentation pipeline for allowing the classifier to learn salient lung features while omitting the outer lung areas of the CT scans
  
  \item Class rebalancing and augmentations, such as intensity- and geometry-based, are employed to develop a general network, although a small dataset is being utilized

\end{itemize}

The remainder of the article is prepared as follows. Section~\ref{materialandMethods} details the materials and methods practiced in the study, including a brief introduction to the methodology and end-to-end 3D-CNN training. Section~\ref{ResultsandDiscussion} describes the experimental operations and their corresponding obtained results. Lastly, section~\ref{Conclusion} concludes the article.

\section{Materials and Methods}
\label{materialandMethods}
In this section, we describe the utilized materials and methods to conduct the widespread experiments. We summarize the adopted dataset in the first subsection~\ref{Dataset}. 
The essential integral preprocessing, such as segmentation, augmentation, and class-rebalancing, are reported in the second subsection~\ref{Preprocessing}.   
The design of the proposed 3D-CNN-based COVID-19 classifier, along with its training protocol, is explained in the third subsection~\ref{Methodologies}.  
Finally, in the fourth subsection~\ref{Hardwareandevaluation}, we represent used hardware to execute the aimed method and evaluation criterion. 

\subsection{Dataset}
\label{Dataset}
This article's experimentations utilize a publicly usable MosMedData dataset administered by municipal hospitals in Moscow, Russia, from March to April 2020 \citep{morozov2020mosmeddata}. 
This dataset includes anonymized human chest lung CT scans with and without COVID-19 related findings of 1110 studies. The population of MosMedData is distributed as $42\,\%$ male, $56\,\%$ female, and $2\,\%$ others, where the median age of the subjects is $47$ years ($18\sim97$ years). All the  studies ($n=1110$) are distributed into five following categories, as presented in Table~\ref{tab:datadistribution}.
\begin{table*}[!ht]
\caption{Distribution of utilized MosMedData dataset for COVID-19 identification  with a short class description.}
\footnotesize
\centering
\begin{tabular}{lp{9cm}ll}
\hline
\rowcolor[HTML]{C0C0C0} 
Class acronym  & Description  & PPI$^*$  & Samples ($\%$) \\ \hline
\textbf{NOR}              & Not consistent with pneumonia, including COVID-19, and refer to a specialist        &  $-$             &     $254$ ($22.8\,\%$)                  \\
 \textbf{MiNCP}              &    Mild novel COVID-19 positive with ground-glass opacities and 
follow-up at home using mandatory telemonitoring          &        $=<25\,\%$         &       $684$ ($61.6\,\%$)                \\
\textbf{MoNCP}              &     Moderate novel COVID-19 positive with ground-glass opacities and follow-up at home by a primary care physician         &        $25-50\,\%$       &       $125$ ($11.3\,\%$)                \\ 
\textbf{SeNCP}              &       Severe novel COVID-19 positive with ground-glass opacities and immediate admission to a COVID specialized hospital        &         $50-75\,\%$       &          $45$ ($4.1\,\%$)             \\ 
\textbf{CrNCP}              &   Critical novel COVID-19 positive with diffuse ground-glass opacities and emergency medical care           &       $>=75\,\%$         &      $2$ ($0.2\,\%$)                 \\ \hline
\multicolumn{3}{l}{Total Samples ($\%$) } &  $1110$ ($100\,\%$)                    \\ \hline
\multicolumn{4}{l}{\scriptsize{PPI$^*$: Pulmonary parenchymal involvement}}             \\ 
\end{tabular}
\label{tab:datadistribution}
\end{table*}
We design two experimental protocols using the MosMedData dataset, such as binary- and multi-class identification, to evaluate our proposed workflow. In binary-class evaluation, we use NOR vs. NCP (Novel COVID-19 Positive), where NCP includes MiNCP-, MoNCP-, SeNCP-, and CrNCP-classes, while in multi-class evaluation, we use NOR vs. MiNCP vs. MoNCP vs. SeNCP.
In multi-class protocols, we merge SeNCP- and CrNCP-classes, naming them as SeNCP, as CrNCP has only two samples in the MosMedData dataset. 
We have applied a cross-validation technique to choose training, validation, and testing images as those are not explicitly given by the data provider. 
The class-wise distribution of MosMedData dataset in Table~\ref{tab:datadistribution} illustrates that the class distribution is imbalanced. Such an imbalanced class distribution produces a biased image classifier towards the class having more training samples. We apply various rebalancing schemes to develop a generic classifier for COVID-19 identification, even though the dataset is imbalanced.

\subsection{Preprocessing}
\label{Preprocessing}
\begin{figure*}[!ht]
  \centering
  \subfloat{\includegraphics[width=16cm, height= 4cm]{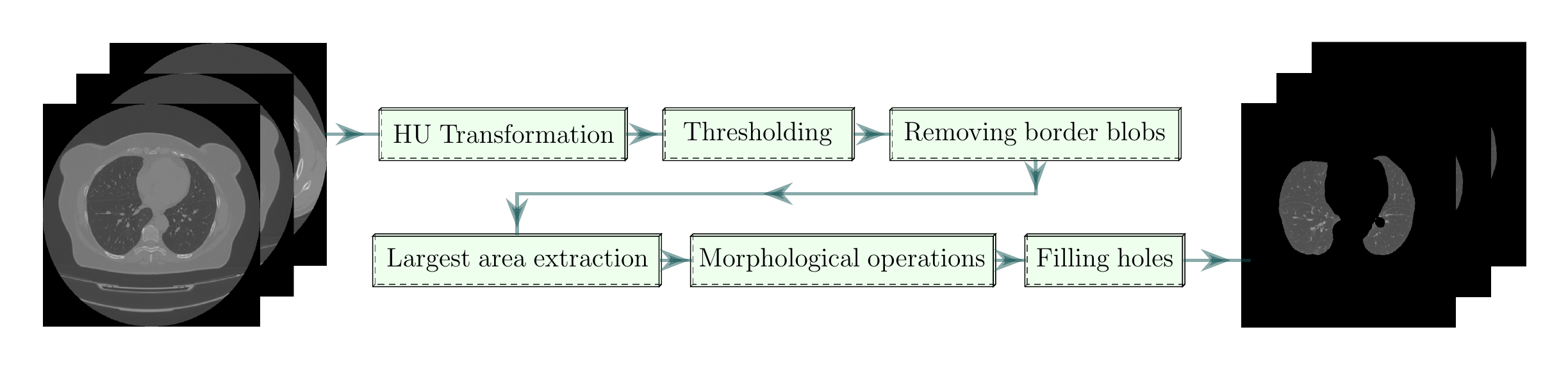} } 
  \caption{The proposed block diagram of an unsupervised lung segmentation pipeline, without requiring a manually annotated lung region.}
  \label{fig:Segmentation}
\end{figure*}

The recommended integral preprocessing consists of segmentation, augmentations (both geometry- and intensity-based), and class-rebalancing, which are concisely explained as follows: 

\textbf{Segmentation.} 
The segmentation, to separate an image into regions with similar properties such as gray level, color, texture, brightness, and contrast, is the significant element for automated detection pipeline \citep{hesamian2019deep}. It is also a fundamental prerequisite for the COVID-19 identification as it extracts the lung region and delivers explanatory information about the shapes, structures, and textures. However, this article proposes an unsupervised Lung Segmentation (LS) technique applying different image processing algorithms, as a massive number of annotated COVID-19 images are not available yet in this pandemic situation. Fig.~\ref{fig:Segmentation} depicts the pipeline of the proposed LS method.  
The proposed threshold-based LS's primary step is transforming all the CT volumes to Hounsfield units (HU), as it is a quantitative measure of radiodensity for CT scans. We set the HU unit as -1000 to -400 as the study shows that lung regions are within that range, which was also used in many articles \citep{shojaii2005automatic, wang2009automated, ko2001chest}. The thresholded binary lung masks are then refined to exclude different false-positive regions, such as the connected blobs with the image border and other small false-positive areas, and false-negative regions, such as small holes in the lung regions. 
Firstly, the border connected regions are eradicated. Secondly, the two largest areas are picked using the region properties algorithm. Thirdly, morphological erosion to separate the lung nodules attached to the blood vessels and morphological closing to keep nodules attached to the lung wall. Finally, the false-negative regions are removed using binary hole fill algorithms.  
Such an unsupervised thresholding-based segmentation method is better in terms of efficiency, taking only a few seconds, and yields utterly reproducible LS.

\textbf{Augmentation.} 
The CNN-based classifiers are profoundly dependent on large data samples to evade the overfitting. Lamentably, various medical imaging fields, especially the current COVID-19 pandemic, suffer from an inadequate dataset size as manually annotated massive training samples are still not available. 
In such a scenario, the augmentations are very dormant preprocessing for increasing the training samples as they are incredibly discriminative \citep{hussain2017differential}. 
Data augmentation incorporates a method that magnifies training datasets' size and property to develop a better-CNN classifier \citep{shorten2019survey}.
The geometric-based augmentation, including a rotation (around $row/2$ and $col/2$) of $-25^{\circ}$, $-15^{\circ}$, $10^{\circ}$, $30^{\circ}$ and height \& width shifting by $20\,\%$, the intensity-based augmentation, including gamma correction \& adding Gaussian random noise, and Elastic deformation\footnote{\url{https://pypi.org/project/elasticdeform/}} are applied in this article as a part of the recommended preprocessing. 
Two values of gamma ($\gamma$), such as $0.7$ and $1.7$, have used in gamma correction to adjust the luminance of the CT volumes by $V_{out}=V_{in}^{\gamma}$, where $V_{out}$ and $V_{in}$ individually denote the output and input values of the luminance.

\textbf{Rebalancing.}
The utilized dataset in Table~\ref{tab:datadistribution} is imbalanced. This situation is pretty obvious in the medical diagnosis field due to the scarcity of massive manually annotated training samples, especially in COVID-19 datasets. The undesired class-biasing occurs in the supervised learning systems towards the class with majority samples.  
However, we apply two techniques to rebalance the imbalanced class distribution, such as adding extra CT volumes from the publicly available CC-CCII dataset \citep{zhang2020clinically} and weighting the loss function for penalizing the overrepresented class. 
The latter approach rewards more extra consideration to the class with minority samples. 
Here, we estimate the class weight using a portion of $W_n = N_n/N$, where $W_n$, $N$, and $N_n$ separately denote the $n^{th}$-class weight, the total sample numbers, and the samples in $n^{th}$-class. 
We employ both the class-rebalancing strategies in the binary-class protocol, whereas the only class weighting method is adopted in the multi-class protocol.

\subsection{Methodologies}
\label{Methodologies}
\subsubsection{Architecture}
\label{Architecture}
The deep neural network is a machine learning framework with a wide range of applications, from natural language processing \citep{deng2018deep} to medical image classification \citep{cai2020review}, segmentation \citep{cai2020review}, and registration \citep{fu2020deep}.
In special, CNNs have become a prevalent technique in the computer vision community. They are practiced in diverse tasks, including object detection \citep{ji2021cnn}, classification \citep{dhruv2020image}, and localization \cite{7827088}. 
The CNN-based deep neural systems are also popularly adopted in recent pandemic for COVID-19 identification \citep{ozsahin2020review,bhattacharya2021deep} (see in Table~\ref{tab:previous_methods_with_results}). 
CNN is an excellent discriminant feature extractor at various abstraction levels, which is translation-invariant. Consequently, utilizing it to classify medical images evades complicated and expensive feature engineering \citep{sarvamangala2021convolutional}. 
The early few CNN layers learn low-level image features and later layers learn high-level image features particular to the application types  \cite{jnawali2018deep}.
However, the 2D-CNNs are frequently employed in natural RGB and grayscale images to extract the spatial features only in two dimensions \citep{szegedy2015going}. The 2D-CNN also can be applied to the volumetric medical image datasets taking cross-sectional 2D slices of the CT, MRI, or similar scans. However, the recent experimental results have revealed the advantages of 3D-CNN over 2D-CNN, where the 3D-CNN accepts the volumetric spatial information as an input \citep{litjens2017survey}. Conventional 2D-CNNs' effectiveness is degraded due to loss of spatial voxel information for volumetric 3D medical imaging tasks. A 3D-CNN, a 3D space implementation of convolution and pooling operation, is practiced to overcome spatial voxel information loss as in the 2D-CNNs. The image becomes scalable in the spatial direction using a 3D-CNN, allowing accurate image detection with different frame sizes \citep{lu20193d}. Therefore, we propose a classifier based on 3D-CNN to identify COVID-19 from the volumetric CT scans.

\begin{figure*}[!ht]
  \centering
\includegraphics[width=17cm, height= 8cm]{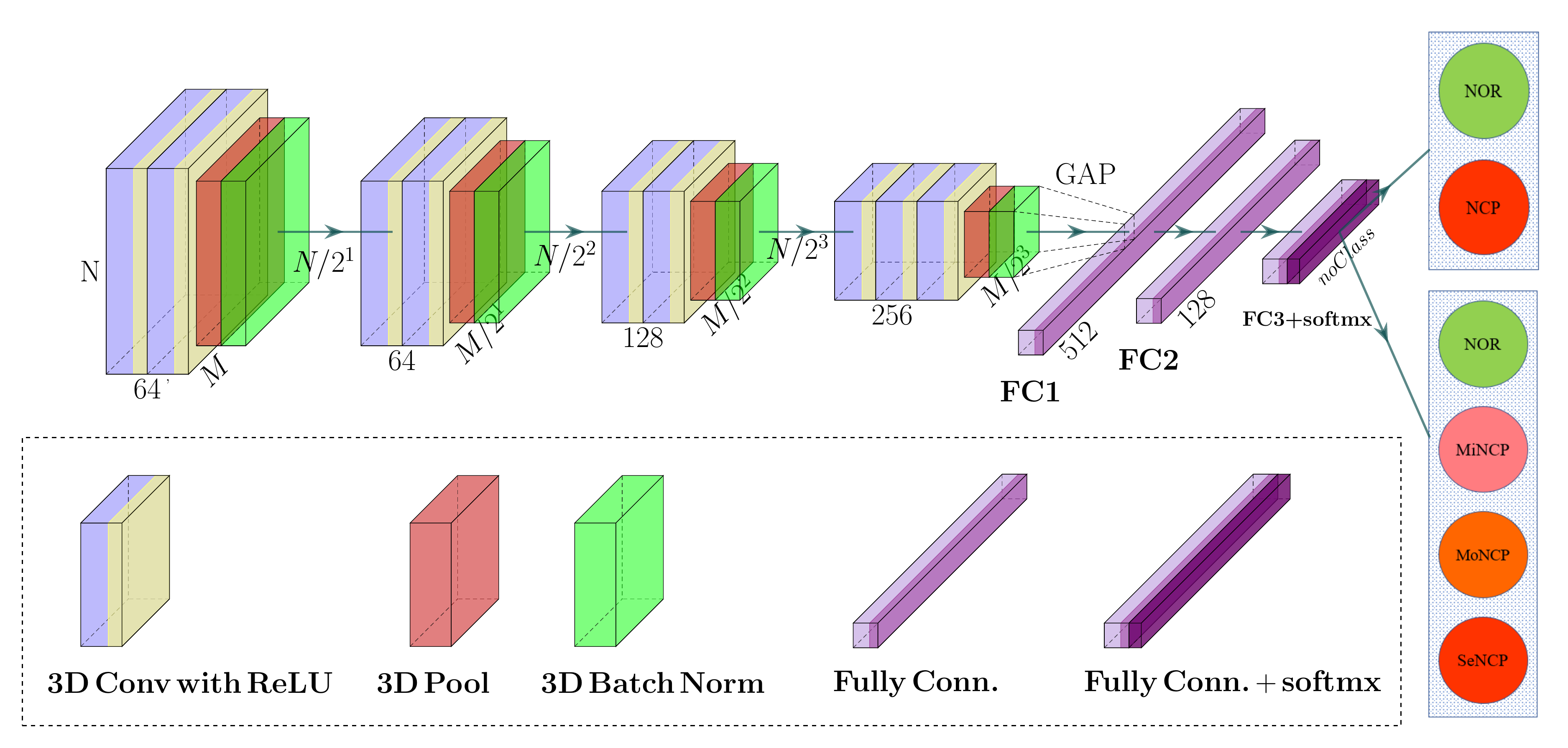}
\caption{The architectural construction of the proposed base network, training with the most smaller 3D patches. This trained base network is applied as a pre-trained model for the next bigger patches. Best view in the color figure.}
\label{fig:base_model}
\end{figure*}

Fig.~\ref{fig:base_model} represents the constructional structure of our proposed COVID-19 base classifier.    
The proposed base network in Fig.~\ref{fig:base_model} essentially consists of two modules, such as feature extractor and feature classifier. The former module is a stack of convolutional, pooling, and batch normalization layers, whereas the latter module is a stack of fully-connected layers followed by a softmax layer. We involve 3D layers for all the feature extractor module components to operate on volumetric medical images for extracting the most discriminating features, accounting for both the intra- and inter-slice spatial voxel information. 
In our network, each 3D convolutional layer with ReLU activation is followed by a 3D max-pooling layer, where the pooling layer increases translational invariances of the network.
The pooled feature maps are then used as an input to the successive layers, which may dynamically change during training at each training epoch \citep{singh20203d}.
The more enormous changes prone to bring difficulties for searching an optimal parameter or hyperparameter; often become computationally expensive to reach an optimal value \citep{ioffe2015batch}. Such a problem is mitigated by integrating batch normalization layers in our network \citep{ioffe2015batch}. It also facilitates the smooth training of the network architectures in less time \citep{singh20203d}.   
The Global Average Pooling (GAP) \citep{lin2013network} is used as a bridge layer between the feature extractor and feature classifier modules, converting the feature tensor into a single long continuous linear vector. 
In GAP, only one feature map is produced for each corresponding category, achieving a more extreme dimensionality compression to evade overfitting \citep{lin2013network}.
A dropout  layer \citep{srivastava2014dropout} is also employed as a regulariser, which randomly sets half of the activation of the fully-connected layers to zero through the training of our network.

\begin{figure*}[!ht]
\centering
\includegraphics[width=17cm, height= 9cm]{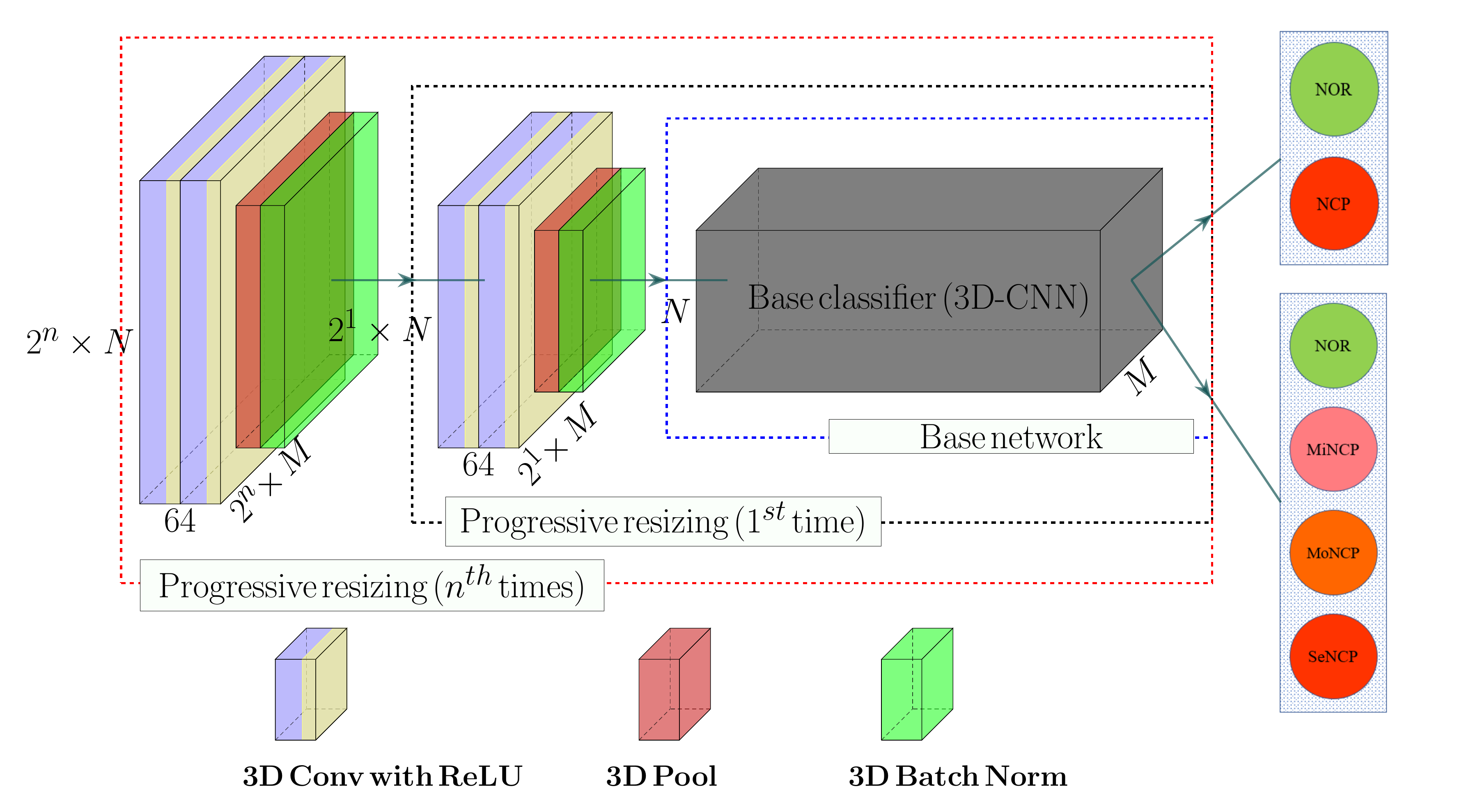}
\caption{The proposed progressively resized network's architectural structure, where the base model (see in Fig.~\ref{fig:base_model}) is trained with the smaller 3D patches and sequentially doubles the base network's size from smaller to larger sizes. The network trained with the smaller patches is the pre-trained model for the next bigger patches. Best view in the color figure.
}
\label{fig:network_PRN}
\end{figure*}

Again, as mentioned earlier, the CNNs are heavily reliant on the massive dataset to bypass overfitting and build a generic network. 
The acquisition of annotated medical images is arduous to accumulate, as the medical data collection and labeling are confronted with data privacy, requiring time-consuming expert explanations \citep{yadav2019deep}. There are two general resolving directions: accumulating more data, such as crowdsourcing \citep{jimenez2018capsule} or digging into the present clinical reports \citep{wang2018tienet}. 
Another technique is investigating how to enhance the achievement of the CNNs with small datasets, which is exceptionally significant because the understanding achieved from the research can migrate the data insufficiency in the medical imaging fields \citep{yadav2019deep}.
Transfer learning is a widely adopted method for advancing the performance of CNNs with inadequate datasets \citep{cheplygina2019not}. 
To our most trustworthy knowledge, there is no public pre-trained 3D-CNN model for the COVID-19 identification from the volumetric chest images with limited samples. 
Therefore, we create a pre-trained model by training our base model (see in Fig.~\ref{fig:base_model}) on the extracted 3D patches from whole chest CT scans (see details in subsection~\ref{Training}). 
Then, we double the patches' size and use them for training the modified base network, where we also double the base model's input size applying a stack of convolutional, pooling, and batch normalization layers (see details in Fig.~\ref{fig:network_PRN}). At the same time, we keep the base model's trained weights for the smaller patches. 
We repeat to enlarge ($n^{th}$-times) the patch and network sizes until we arrive at the provided CT scans' size, as pictured in Fig.~\ref{fig:network_PRN}.  
Such training is called progressive resizing \cite{arani2021rgpnet}, where the training begins with smaller image sizes followed by a progressive expansion of size. This training process is continued until the last patch and network sizes are as same as the initial image dimension.

\subsubsection{Training protocol}
\label{Training}
We first extract five different patches with different sizes (see in Fig.~\ref{fig:Patches}) to begin the experimentations. We perform ablation studies in subsection~\ref{PatchSelection} looking for the best patch size. The weights of the base network in Fig.~\ref{fig:base_model} is initialized with Xavier normal distribution. The weights of the first progressively resized network are initialized with the weights of the base network.  In general, the weights of the network with the patch size of $(R/2^{n})\times (C/2^{n}) \times (S/2^{n})$ are initialized with the weights of the network with the patch size of $(R/2^{n+1})\times (C/2^{n+1}) \times (S/2^{n+1})$ for the original CT volume size of $R\times C \times S$.  

Categorical cross-entropy and accuracy are utilized as a loss function and metric, respectively, for training all the networks in this article. We use Adam \citep{kingma2014adam} optimizer with initial learning rate ($LR$), exponential decay rates ($\beta{_1},\,\beta{_2}$) as $LR=0.0001$, $\beta{_1}=0.9$, and $\beta{_2}=0.999$, respectively, without AMSGrad variant. The exponential decaying LR schedule is also employed for the networks' optimization. Initial epochs are set as 200, and training is terminated if validation performance stops growing after 15 epochs.

\subsection{Hardware and evaluation criterion}
\label{Hardwareandevaluation}
We execute all the comprehensive experiments on a \textit{Windows-$10$} machine utilizing the Python, with various Keras \citep{geron2019hands} and image processing APIs, and MATLAB programming languages. 
The device configurations of the used machine are: 
Intel\textsuperscript{\tiny\textregistered} Core\textsuperscript{\tiny{TM}} i$7$-$7700$ HQ CPU @ $3.60\,GHz$ processor with a install memory (RAM) of $32.0\,GB$, and GeForce GTX $1080$ GPU with a memory of $8.0\,GB$ (GDDR$5$).

We evaluate all the experimental outcomes by employing numerous metrics, such as recall, precision, and F1-score, for evaluating them from diverse perspectives.
The recall measures the type-II error (the patient having positive COVID-19 characteristics, erroneously abandons to be repealed), whereas the precision estimates the positive predictive values (a portion of absolutely positive-identification amid all the positive-identification). 
The harmonic mean of recall and precision is manifested using the F$1$-score, conferring the tradeoff between these two metrics.
Furthermore, we also quantify the prognostication probability of an anonymously selected CT sample using a Receiver Operating Characteristics (ROC) with its Area Under the ROC Curve (AUC) value.

\section{Results and Discussion}
\label{ResultsandDiscussion}
In this section, the achieved results from different experiments are reported with comprehensive discussion. 
In subsection~\ref{PatchSelection}, we confer the results of COVID-19 identification utilizing various 3D patches and compare them with original CT image utilization on the same experimental conditions and network. We discuss the results of progressive resizing over a single fixed size in subsection~\ref{ProgressiveResizing}. We demonstrate the effects of different proposed preprocessing on COVID-19 identification in subsection~\ref{PrepossessingEmployment}. Finally, in subsection~\ref{BinaryVsMulticlassEvaluation}, we show the results for binary- and multi-class COVID-19 identification applying our proposed network and preprocessing.

\subsection{Patch Selection}
\label{PatchSelection}
\begin{figure*}[!ht]
  \centering
\includegraphics[width=13cm, height= 9cm]{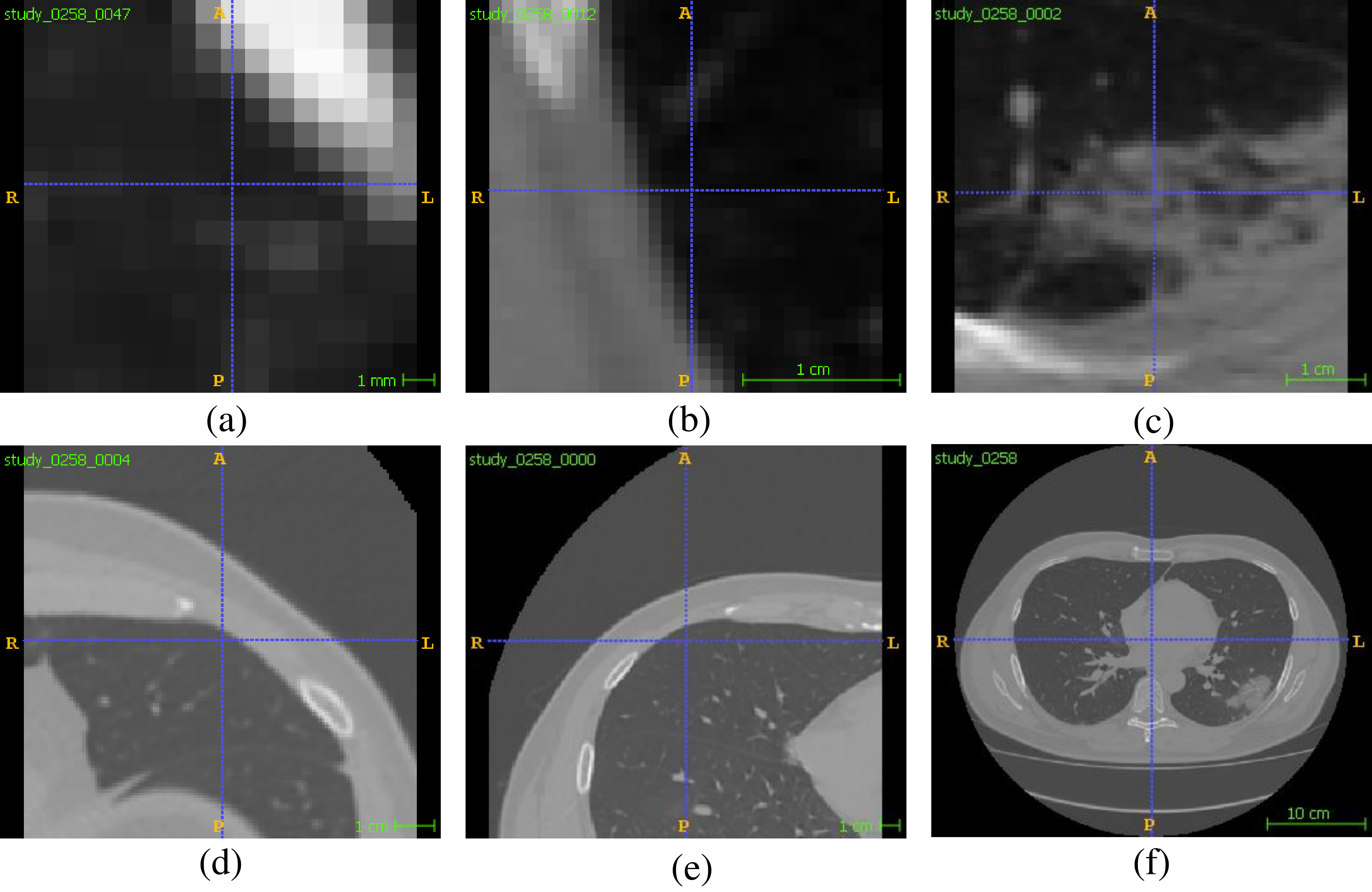}
\caption[]{Example of various extracted patches having different sizes, as mentioned earlier, where patches $P_1$ to $P_6$ are displayed in a) to f), respectively. The middle slices of each 3D patches are illustrated for the same sample (\textit{study\_0258.nii.gz}) in the MosMedData dataset. Slices are captured using a ITK-Snap windows version\footnote{\url{http://www.itksnap.org/pmwiki/pmwiki.php?n=Downloads.SNAP3}}.}
\label{fig:Patches}
\end{figure*}

We extract five different 3D patches, named $P_1$, $P_2$, $P_3$, $P_4$, and $P_5$, having respective size of $16\times 16\times 9$, $32\times 32\times 12$, $64\times 64\times 15$, $128\times 128\times 20$, and $256\times 256\times 27$. The original CT scans having size of $512\times 512\times 36$ is named as $P_6$. The height and width of the patch $P_5$ is half of the $P_6$, whereas these dimensions of the patch $P_4$ is one-fourth of the $P_6$, and so on. We extract $2^n$ number of patches for a $n^{th}$-time reduction of the height and width. Therefore, we train and test our network with $71040$, $35520$, $17760$, $8880$, $4440$, and $1110$ samples for the 3D volumes $P_1$ to $P_6$, respectively. The examples of the extracted patches are shown in Fig.~\ref{fig:Patches}, where we select the middle slices of the extracted patches of the same CT scan.    
Different patches in Fig.~\ref{fig:Patches} shows their respective resolutions, where it is seen that the patches $P_1$ and $P_2$ demonstrate very low resolutions. However, the effects of those patch resolutions are judged by classifying the NOR vs. NCP classes (see in subsection~\ref{Dataset}).

\begin{figure*}[!ht]
  \centering
\includegraphics[width=16cm, height= 8cm]{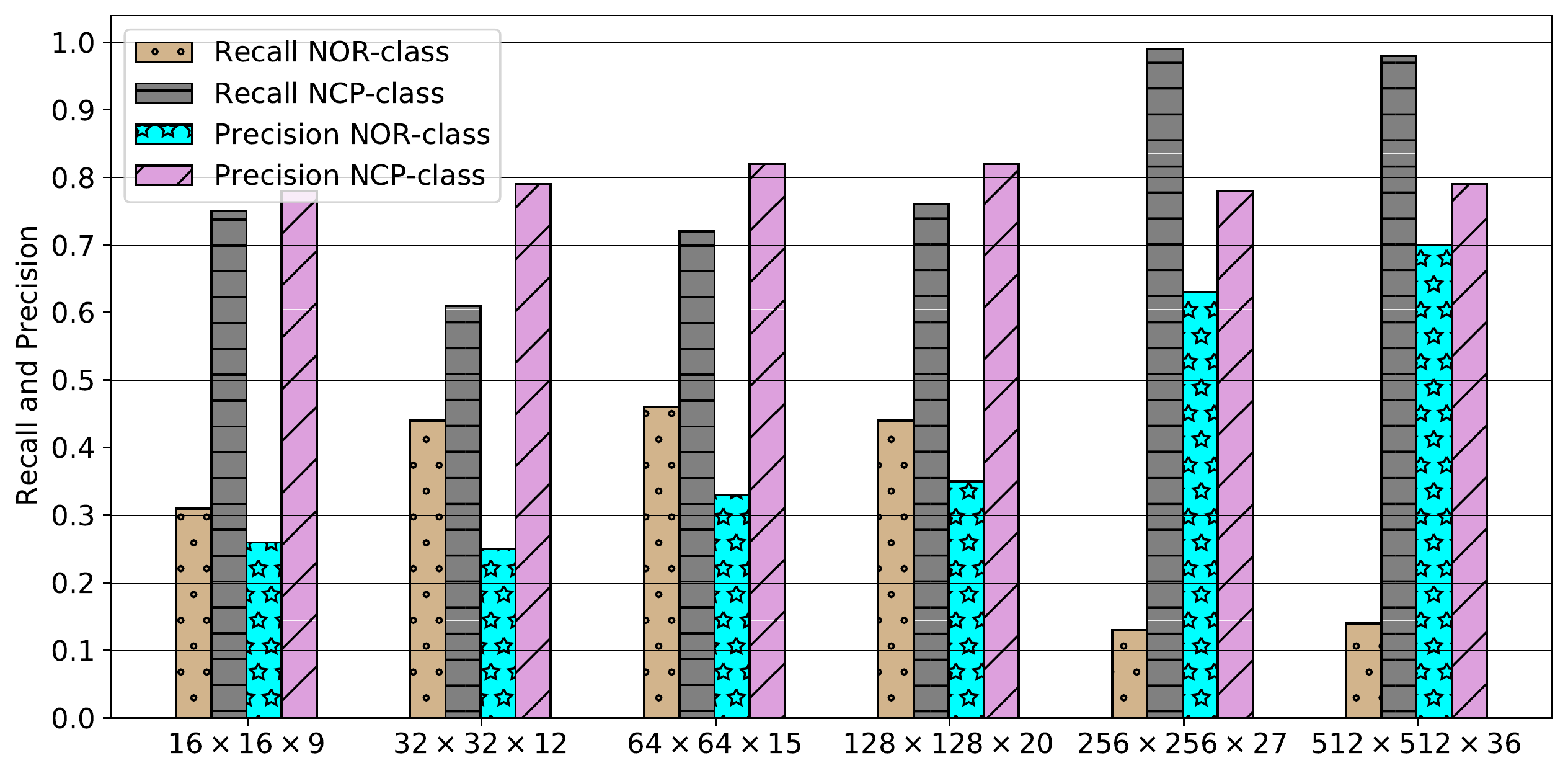}
\caption{The binary classification results from our 3D-CNN utilizing different 3D-patch sizes, where the bars with dots, horizontal lines, stars, and diagonal hatching respectively denote recall and precision of NOR- and NCP-classes. Best view in the color figure.}
  \label{fig:Patch_Selection}
\end{figure*}

The classification results are presented in Fig.~\ref{fig:Patch_Selection} for all the patches ($P_1$ to $P_5$) and original CT scans ($P_6$) employing our 3D network without any type of preprocessing.  
The results show that the network inputting with $P_1$ patch outputs COVID-19 identification with type-II errors as $69.0\,\%$ and $25.0\,\%$ for NOR- and NCP-classes, respectively. Such results confirm that NCP-class has been identified more accurately ($44.0\,\%$ more in NCP-class), pointing that classifier is biased towards the NCP-class.
On the other hand, the utilization of patch $P_2$ produces identification results with type-II errors as $56.0\,\%$ and $39.0\,\%$ for NOR- and NCP-classes, which reduce the differences between two classes (only $17.0\,\%$ more in NCP-class). Although the $P_1$ patch has double samples, it fails to provide a class-balanced performance than the $P_2$ patch. This is because of having a better-resolution in the $P_2$ patch than the $P_1$ patch (see in Fig.~\ref{fig:Patches}), as other experimental settings are constant. 
Again, the patch $P_3$ further improves the identification results with type-II errors as $54.0\,\%$ and $28.0\,\%$ for NOR- and NCP-classes.
Approximately, the patch $P_4$ also provides similar results to the $P_3$ patch.
It is noteworthy from those experimentations that $P_3$ or $P_4$ patches have much fewer samples than $P_1$ ($4$-times and $8$-times, respectively); still, they outperform the identification results of $P_1$ and $P_2$ patches with the same experimental settings.    

Furthermore, the utilization of patch $P_5$ further reduces the performances (type-II errors as $6.0\,\%$ and $99.0\,\%$ for NOR- and NCP-classes) than all the previous patches discussed above. Such a result shows that it produces a more biased model towards the NCP-class. 
From Fig.~\ref{fig:Patches} shows that the patch $P_4$ and $P_5$ are visually looking similar but $P_4$ has two-times samples as of $P_5$.
This experiment exposes that having fewer samples also generates class-biased classifiers if input images are similar in resolution.  

Finally, the network with the original images also provides less COVID-19 identification performance as in the patch $P_5$ (see in Fig.~\ref{fig:Patch_Selection}). All the experiments show that our network with $P_3$ or $P_4$ patches has outputted better-identification results. Such experimental results undoubtedly prove that both the input resolution and the number of samples play an important role in CNN-based classifiers. We can not increase the number of samples taking the smaller patch sizes, as it has a shallow resolution, which adversely affects the classifiers.

\subsection{Progressive Resizing}
\label{ProgressiveResizing}
The aforementioned results reveal that the utilization of better-resolution with more sample numbers increases the performance of CNN. Therefore, we propose to employ progressive resizing of our proposed 3D-CNN (see details in subsection~\ref{Methodologies}). Firstly, we begin training our network with a suitable 3D patch with more training samples from the previous experiments, acting as a base model. Then, we add some CNN layers to the input of the base model with the higher resolution ($2$-times more in this article), where the base model is adopted as a pre-trained model (see details in subsection~\ref{Methodologies}). We repeat this network resizing until we reach to original given CT size ($P_6$).  

\begin{figure*}[!ht]
  \centering
\subfloat{\includegraphics[width=8.4cm, height= 6.5cm]{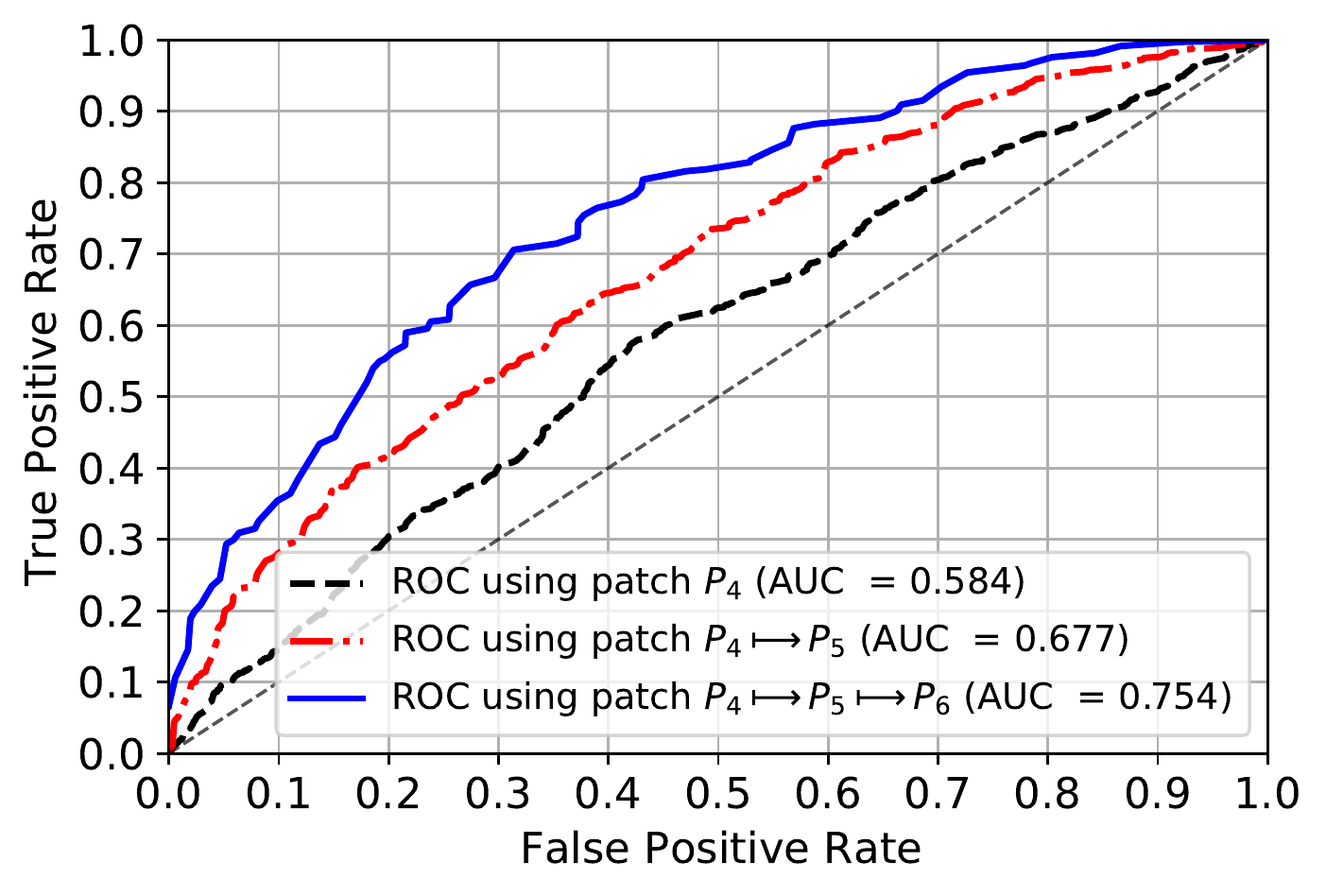}}
\caption{The ROC curves for the progressive resizing of our 3D network. Best view in the color figure.}
 \label{fig:ROC_Progressive_resizing}
\end{figure*}

The results for such a progressive resizing are presented in the confusion metrics in Table~\ref{tab:confusion} and ROC curves (with respective AUC values) in Fig.~\ref{fig:ROC_Progressive_resizing}. 
\begin{table*}[!ht]
\footnotesize
\centering
\caption{Normalized confusion matrix employing our network with progressive resizing, where we progressively increase the input resolution from $P_4$ to $P_5$ then to $P_6$ (original resolution). The first table (left) for the resolution of $P_4$, the second table (middle) for resolution of $P_4 \longmapsto P_5$, and the last (right) for resolution of $P_4 \longmapsto P_5 \longmapsto P_6$.}
\begin{tabular}{lll}

\begin{tabular}{cccc}
\hline
\multicolumn{2}{c}{}                         & \multicolumn{2}{c}{Actual}                            \\ \cline{3-4} 
\multicolumn{2}{c}{\multirow{-2}{*}{$P_4$}} & NOR                       & NCP                       \\ \hline
                                  & NOR      & \cellcolor[HTML]{C0C0C0}$24.26\,\%$ &                 $13.52\,\%$          \\
\multirow{-2}{*}{\rotatebox[origin=c]{90}{Predict}}       & NCP      &                         $75.74\,\%$  & \cellcolor[HTML]{C0C0C0}$86.48\,\%$ \\ \hline
\end{tabular}

&

\begin{tabular}{cccc}
\hline
\multicolumn{2}{c}{}                         & \multicolumn{2}{c}{Actual}                            \\ \cline{3-4} 
\multicolumn{2}{c}{\multirow{-2}{*}{$P_4 \longmapsto P_5 $}} & NOR                       & NCP                       \\ \hline
                                  & NOR      & \cellcolor[HTML]{C0C0C0}$21.08\,\%$ &                  $5.38\,\%$         \\
\multirow{-2}{*}{\rotatebox[origin=c]{90}{Predict}}       & NCP      &  $78.92\,\%$                         & \cellcolor[HTML]{C0C0C0}$94.62\,\%$ \\ \hline
\end{tabular}

&

\begin{tabular}{cccc}
\hline
\multicolumn{2}{c}{}                                                                      & \multicolumn{2}{c}{Actual}                            \\ \cline{3-4} 
\multicolumn{2}{c}{\multirow{-2}{*}{\begin{tabular}[c]{@{}c@{}}$P_4 \longmapsto P_5 $\\ $P_5 \longmapsto P_6$ \end{tabular}}} & NOR                       & NCP                       \\ \hline
                                                        & NOR                             & \cellcolor[HTML]{C0C0C0}$39.22\,\%$  &   $9.30\,\%$                         \\
\multirow{-2}{*}{\rotatebox[origin=c]{90}{Predict}}                             & NCP                             &          $60.78\,\%$                  & \cellcolor[HTML]{C0C0C0}$90.70\,\%$  \\ \hline
\end{tabular}
\end{tabular}
\label{tab:confusion}
\end{table*}
The confusion matrix in Table~\ref{tab:confusion}, for more detailed analysis of the identification results, points that $24.26\,\%$-NOR samples are accurately classified as NOR, whereas $86.48\,\%$-NCP samples are correctly classified as NCP while utilizing the 3D patch $P_4$ with $8880$ samples. This training is set as a base model. 
Now, employing the base model as a pre-trained model, the utilization of $P_5$ patch, with $4440$ samples, decreases the false-negative rate of NCP by $8.14\,\%$ although false-positive rate increases by $3.18\,\%$ (see in Table~\ref{tab:confusion} (left and middle)). This training is a first-time progressive resizing ($P_4 \longmapsto P_5 $).  
Again, employing $P_4 \longmapsto P_5 $ as a pre-trained model, the utilization of $P_6$ (original CT scans), with $1110$ samples, increases the false-negative rate of NCP by $3.92\,\%$, still less than baseline false-negative rate of $13.52\,\%$. It also decreases the false-positive rate by a margin of $18.14\,\%$, which is less than the former two false-positive rates (see all tables in Table~\ref{tab:confusion}).
Furthermore, the proposed final progressively resized network ($P_4 \longmapsto P_5 \longmapsto P_6$) obtains an AUC of $0.754$, which indicates that the probability of correct COVID-19 identification is as high as $75.4\,\%$ for any given random CT samples (see in Fig.~\ref{fig:ROC_Progressive_resizing}). 
It has beaten the baseline  $P_4$ and $P_4 \longmapsto P_5$ respectively by $17.0\,\%$ and $7.70\,\%$ in terms of AUC, as presented in Fig.~\ref{fig:ROC_Progressive_resizing}.
Although the final progressively resized network ($P_4 \longmapsto P_5 \longmapsto P_6$) has an input of the original CT scans, its performance is far better than the network training with $P_6$ alone (see in Fig.~\ref{fig:Patch_Selection}).
All the above discussions in this subsection experimentally validate the progressive resizing supremacy for the COVID-19 identification instead of training using single size input CT scans.

\subsection{Prepossessing Employment}
\label{PrepossessingEmployment}
This subsection presents the COVID-19 identification results from our progressively resized 3D network employing different preprocessing, such as augmentation, segmentation, and class-rebalancing. 

Table~\ref{tab:different_exp} bestows different experimental results, where we explicitly explicate the outcomes of each preprocessing for the  COVID-19 identification from volumetric CT scans. 
\begin{table*}[!ht]
\caption{The COVID-19 identification results on the MosMedData dataset from our 3D-CNN network utilizing different preprocessing.}
\footnotesize
\centering
\begin{tabular}{lcccccc}
\hline
\rowcolor[HTML]{C0C0C0} 
\cellcolor[HTML]{C0C0C0}                                        & \multicolumn{6}{c}{\cellcolor[HTML]{C0C0C0}Class-wise and weighted average metrics}                        \\ \cline{2-7} 
\rowcolor[HTML]{C0C0C0} 
\cellcolor[HTML]{C0C0C0}                                        & \multicolumn{3}{c}{\cellcolor[HTML]{C0C0C0}Recall} & \multicolumn{3}{c}{\cellcolor[HTML]{C0C0C0}Precision} \\ \cline{2-7} 
\rowcolor[HTML]{C0C0C0} 
\multirow{-3}{*}{\cellcolor[HTML]{C0C0C0}Different experiments} & NOR             & NCP            & Avg.            & NOR              & NCP             & Avg.             \\ \hline

                    Baseline model       &    $0.137$             &     $0.983$           &    $0.789$             &       $0.700$           &     $0.793$            & $0.772$ \\

                                      Progressively Resized Network (PRN)              &    $0.392$             &      $0.907$          &  $0.789$               &    $0.556$              &     $0.834$            &           $0.770$       \\
                                                         PRN with Augmentation (PRNA)       &                 $0.529$             &      $0.884$          &  $0.803$               &    $0.574$              &     $0.864$            &           $0.798$              \\
                                              PRN with Lung Segmentation (PRNS)                  &  $0.333$               &  $0.971$              &    $0.825$             &     $0.773$             &  $0.831$               & $0.818$ \\ 
                PRNA and PRNS with Class-rebalancing (PRNASCR)       &  $0.706$               &  $0.919$              &  $0.870$               &     $0.720$             &   $0.913$              & $0.869$ \\ \hline
\end{tabular}
\label{tab:different_exp}
\end{table*}
The baseline model, without progressive resizing and inputting with original CT scans ($P_6$), produces low identification consequences resulting in type-II errors of $86.3\,\%$ and $1.7\,\%$ respectively for NOR- and NCP-classes, showing high class-imbalanced results. The weighted average type-II error is also only $21.1\,\%$ with respective average positive predictive value as $77.2\,\%$. 
Highly imbalanced training samples ($NOR:NCP=1:3.37$) with less intra-class heterogeneity and high inter-class similarity are the probable causes for providing such a poor result. 
However, the utilization of different 3D patches improves intra-class heterogeneity and inter-class similarity and appliance of progressive resizing, where the base model acts as a pre-trained model, can mitigate those aforementioned difficulties, which reflects in the PRN results (see in the second row of Table~\ref{tab:different_exp}).    
The appliance of PRN successfully reduces the class-imbalanced results improving the type-II error of NOR-class by a margin of $25.5\,\%$, while the weighted average type-II error is identical ($21.1\,\%$).  

\paragraph{\textbf{Augmentation}} The employment of different image augmentations, such as random rotation, height \& width shifting, gamma correction, adding Gaussian noise, and Elastic deformation (see details in subsection~\ref{Preprocessing}) with PRN further improves the COVID-19 identification results, showing far better class-balance (type-II error of NOR-class improved by a margin of $13.7\,\%$ with significantly less reduction as $2.3\,\%$ in NCP-class). The weighted average type-II error is increased by $1.4\,\%$ with respective increases in average positive predictive value by $2.8\,\%$ for the appliance of augmentations with the PRN.   

\begin{figure*}[!ht]
\centering
\subfloat{\includegraphics[width=16.4cm, height= 5cm]{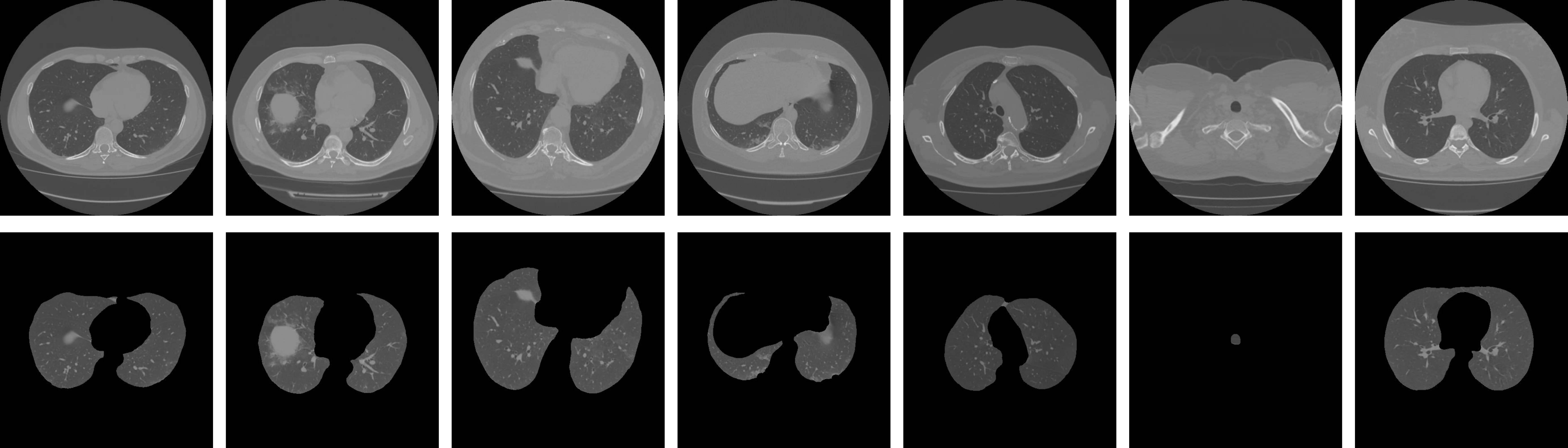}}
\caption{Examples of lung segmentation results applying our unsupervised pipeline, as described in subsection~\ref{Preprocessing}.}
 \label{fig:lungSegmentation}
\end{figure*}

\paragraph{\textbf{Segmentation}}
The well-defined segmentation, with less-coarseness, is an essential requirement for further identification. 
The incorporation of segmentation with the PRN further promotes the identification results than the PRN alone, as exposed in Table~\ref{tab:different_exp}. Several examples of the segmented lung from our proposed unsupervised pipeline (as described in subsection~\ref{Preprocessing}) are depicted in Fig.~\ref{fig:lungSegmentation} for qualitative evaluation.  
However, the COVID-19 identification results incorporating lung segmentation with the PRN reflects its supremacy over the PRN alone, extending the weighted average type-II error by $1.4\,\%$ with respective improvements in average positive predictive value by $2.8\,\%$ (see in Table~\ref{tab:different_exp}). The class-imbalanced identification is also dwindled due to segmented lung area utilization over the full CT volumes.
The reasonable ground for those enhanced performances due to the segmentation is that it extracts an abstract region, enabling the classifier to learn only the precise lung areas' features while avoiding the surrounding healthy tissues of the chest CT scans.

\begin{figure*}[!ht]
\centering
\subfloat{\includegraphics[width=8.4cm, height= 6.5cm]{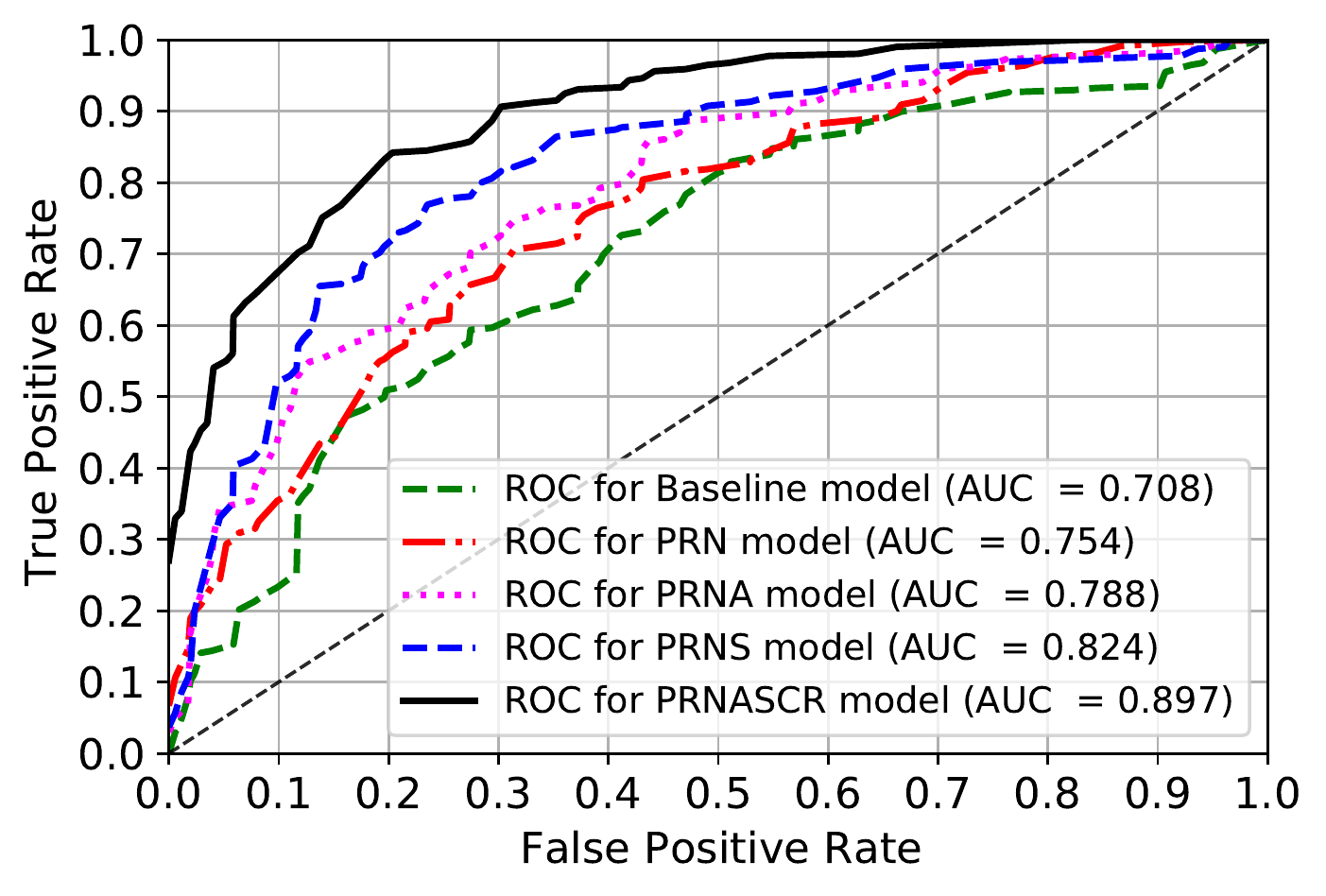}}
\caption{The ROC curves for the employment of various preprocessing to our 3D network. Best view in the color figure.}
\label{fig:ROC_all_expts_preprocessing}
\end{figure*}

\paragraph{\textbf{Augmentation, Segmentation, and Class-rebalancing}} 
The combination of augmentations, segmentation, and class-rebalancing with the PRN provides the best COVID-19 identification results of this article. This experiment identifies the COVID-19 from the chest CT scans with relatively less class-imbalance with the weighted average type-II error of $13.0\,\%$ with respective average positive predictive value as $13.1\,\%$. All the preprocessing employment heightens the former metric by a margins of $8.1\,\%$ and the latter metric by $9.7\,\%$ from the baseline model (see in Table~\ref{tab:different_exp}) with less class-imbalance performance. 
Besides, Fig.~\ref{fig:ROC_all_expts_preprocessing} displays the ROC curves of our PRN with/without all the preprocessing and a baseline model with their corresponding AUC values. 
The proposed PRNASCR achieves an AUC of $0.897$, showing the probability of accurate COVID-19 recognition is as large as $89.7\,\%$ for any yielded random CT sample. For AUC, the proposed PRNASCR betters the baseline model, PRN, PRNA, and PRNS respectively by $18.9\,\%$, $14.3\,\%$, $10.9\,\%$, and $7.3\,\%$. 
From Fig.~\ref{fig:ROC_all_expts_preprocessing} and given $10.0\,\%$ false-positive rates, the true-positive rates of COVID-19 identification from the baseline model, PRN, PRNA, PRNS, and PRNASCR are approximately $22.0\,\%$, $35.0\,\%$, $46.0\,\%$, $52.0\,\%$, and $67.0\,\%$, respectively, showing the improvements of $45.0\,\%$ from the baseline $22.0\,\%$.

\subsection{Binary- Vs. Multi-class Evaluation}
\label{BinaryVsMulticlassEvaluation}
This subsection displays the COVID-19 identification results using our proposed PRNASCR for binary- and multi-class (see in subsection~\ref{Dataset}) utilizing the 5-fold cross-validation. 
\begin{table*}[!ht]
\footnotesize
\centering
\caption{The confusion matrix for the COVID-19 identification on the MosMedData dataset from our proposed 3D-CNN network and preprocessing for both the binary- (left) and multi- (right) class problems.}
\begin{tabular}{ll}
\begin{tabular}{cccc}
\hline
\multicolumn{2}{c}{}                         & \multicolumn{2}{c}{Actual}                            \\ \cline{3-4} 
\multicolumn{2}{c}{\multirow{-2}{*}{2-classes}} & NOR                       & NCP                       \\ \hline
                                  & NOR      & \cellcolor[HTML]{C0C0C0}\begin{tabular}[c]{@{}c@{}}167\\ $65.75\,\%$\end{tabular}&                 \begin{tabular}[c]{@{}c@{}}8\\ $0.94\,\%$\end{tabular}          \\
\multirow{-3}{*}{\rotatebox[origin=c]{90}{Predict}}       & NCP      &                         \begin{tabular}[c]{@{}c@{}}87\\ $34.25\,\%$\end{tabular}  & \cellcolor[HTML]{C0C0C0}\begin{tabular}[c]{@{}c@{}}848\\ $99.06\,\%$\end{tabular} \\ \hline
\end{tabular}

&

\begin{tabular}{cccccc}
\hline
\multicolumn{2}{c}{}                            & \multicolumn{4}{c}{Actual}        \\ \cline{3-6} 
\multicolumn{2}{c}{\multirow{-2}{*}{4-classes}} & NOR                                                                          & MiNCP                                                                        & MoNCP                                                                        & SeNCP                                                                        \\ \hline
                                 & NOR          & \cellcolor[HTML]{C0C0C0}\begin{tabular}[c]{@{}c@{}}188\\ $74.02\,\%$\end{tabular} & \begin{tabular}[c]{@{}c@{}}67\\ $9.80\,\%$\end{tabular}                         & \begin{tabular}[c]{@{}c@{}}3\\ $2.40\,\%$\end{tabular}                         & \begin{tabular}[c]{@{}c@{}}2\\ $4.26\,\%$\end{tabular}                         \\
                                 & MiNCP        & \begin{tabular}[c]{@{}c@{}}62\\ $24.41\,\%$\end{tabular}                         & \cellcolor[HTML]{C0C0C0}\begin{tabular}[c]{@{}c@{}}580\\ $84.80\,\%$\end{tabular} & \begin{tabular}[c]{@{}c@{}}29\\ $23.2\,\%$\end{tabular}                         & \begin{tabular}[c]{@{}c@{}}13\\ $27.66\,\%$\end{tabular}                         \\
                                 & MoNCP        & \begin{tabular}[c]{@{}c@{}}3\\ $1.18\,\%$\end{tabular}                         & \begin{tabular}[c]{@{}c@{}}22\\ $3.22\,\%$\end{tabular}                         & \cellcolor[HTML]{C0C0C0}\begin{tabular}[c]{@{}c@{}}86\\ $68.80\,\%$\end{tabular} & \begin{tabular}[c]{@{}c@{}}1\\ $2.13\,\%$\end{tabular}                         \\
\multirow{-7}{*}{\rotatebox[origin=c]{90}{Predict}}        & SeNCP        & \begin{tabular}[c]{@{}c@{}}1\\ $0.39\,\%$\end{tabular}                         & \begin{tabular}[c]{@{}c@{}}15\\ $2.18\,\%$\end{tabular}                         & \begin{tabular}[c]{@{}c@{}}7\\ $5.60\,\%$\end{tabular}                         & \cellcolor[HTML]{C0C0C0}\begin{tabular}[c]{@{}c@{}}31\\ $65.95\,\%$\end{tabular} \\ \hline
\end{tabular}
\end{tabular}
\label{tab:confusion_folds}
\end{table*}
The detailed class-wise performance of our PRNASCR for both the binary- and multi-class is exhibited in the confusion metrics in 
Table~\ref{tab:confusion_folds} (left) and Table~\ref{tab:confusion_folds} (right), correspondingly.

\begin{figure*}[!ht]
  \centering
\subfloat[ROC for binary-class]{\includegraphics[width=8.4cm, height= 6cm]{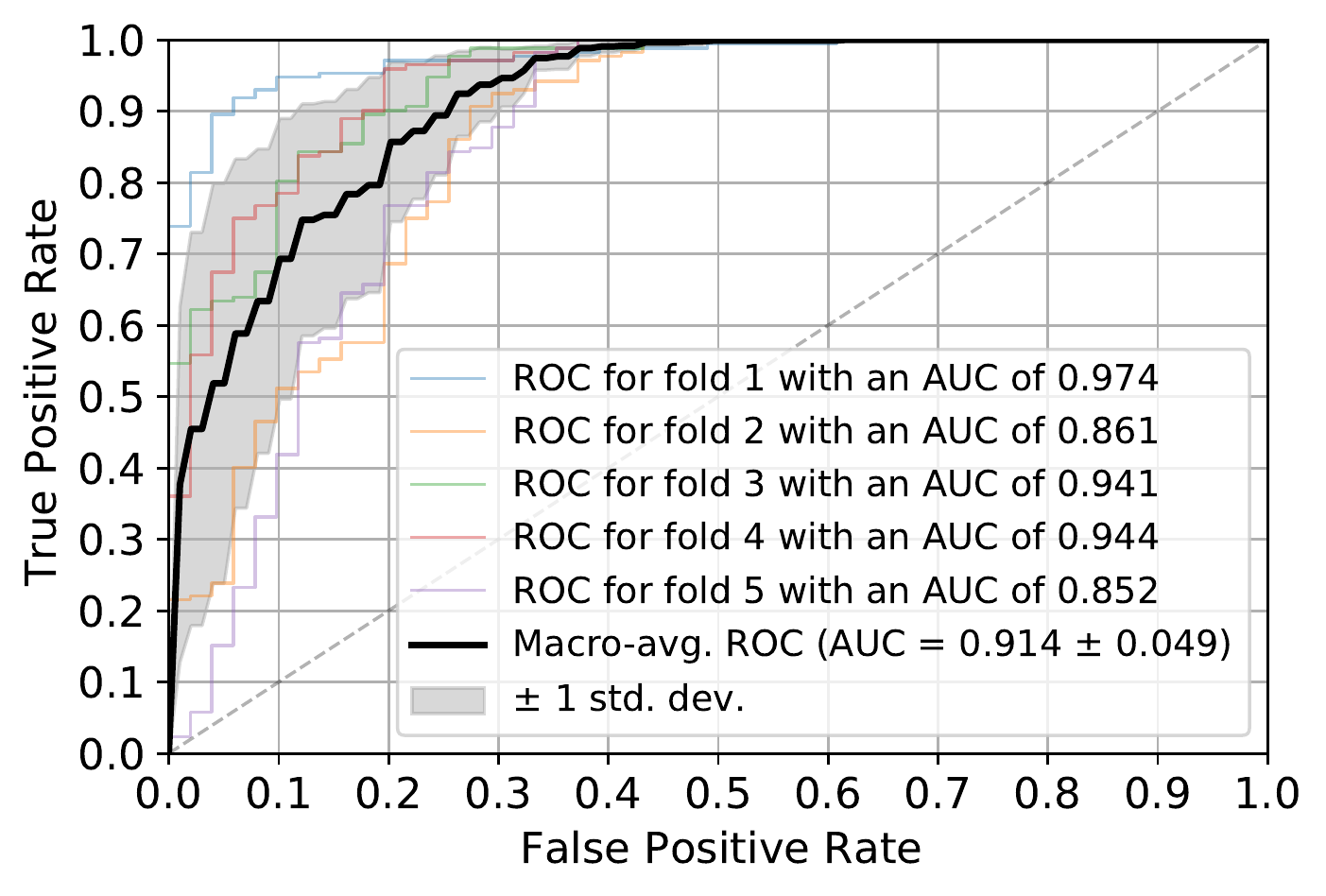}}
\subfloat[ROC for multi-class]{\includegraphics[width=8.4cm, height= 6cm]{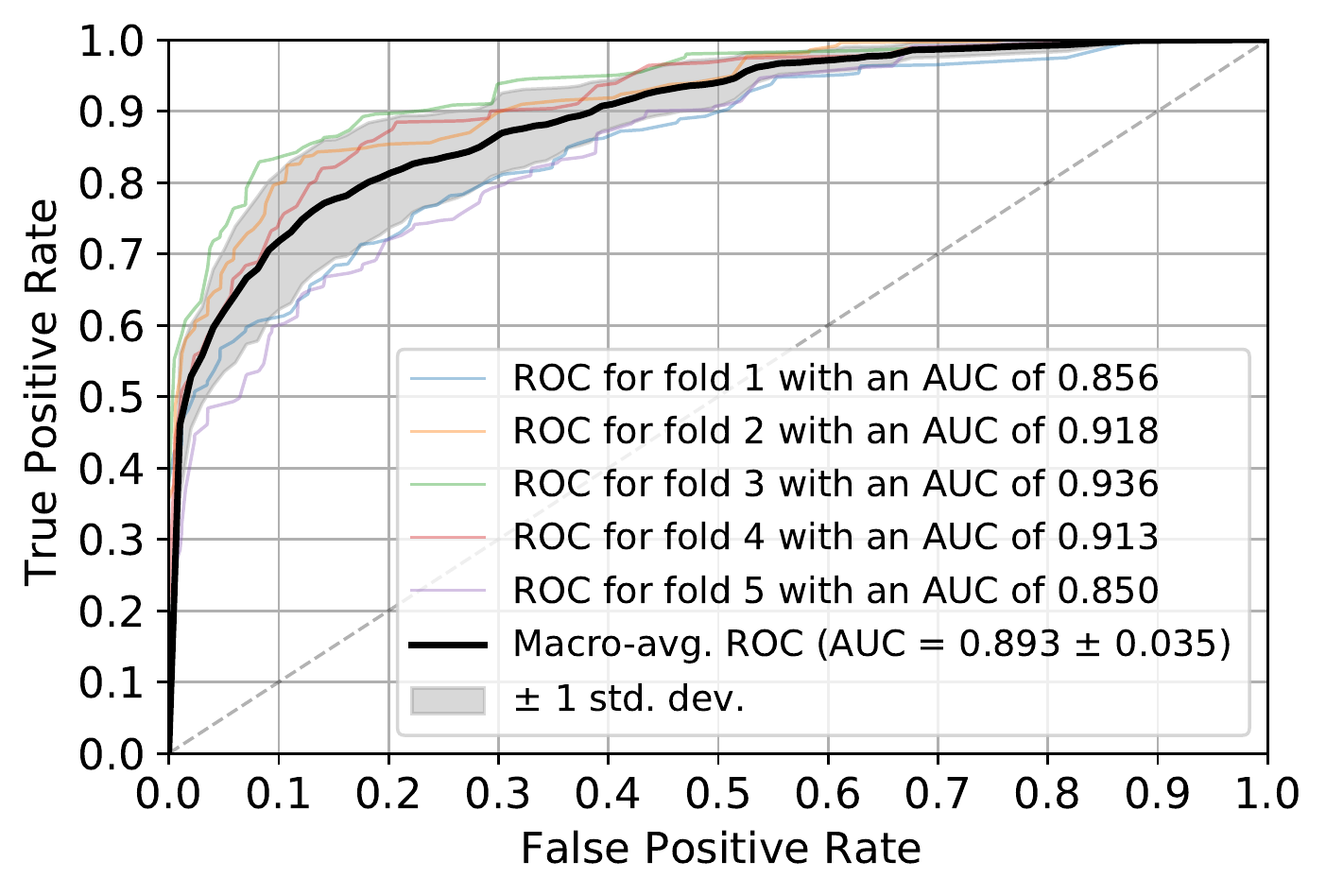}}
\caption{The ROC curves for the binary- and multi-class identification of COVID-19, applying 5-fold cross-validations. Best view in the color figure.}
 \label{fig:ROC_all_expts_folds}
\end{figure*}

The binary-classification results in Table~\ref{tab:confusion_folds} (left) show that among $254$-NOR CT samples, correctly classified samples are $167\,(67.75\,\%)$, whereas only $87\,(34.25\,\%)$ samples are erroneously classified as NCP (false positive).
It is also noteworthy that among $856$-NCP samples, rightly classified samples are $848\,(99.06\,\%)$, whereas only $8\,(0.94\,\%)$ samples are wrongly classified as NOR (false negative). 
Again, the matrix in Table~\ref{tab:confusion_folds} (right) for multi-class recognition reveals the FN and FP for the COVID-19 identification, where number of wrongly classified CT images (type-I or type-II errors) are $66/256\,(25.78\,\%)$, $104/684\,(15.20\,\%)$, $39/125\,(31.20\,\%)$, and $16/47\,(34.04\,\%)$ respectively for the NOR-, MiNCP-, MoNCP-, and SeNCP-classes.
Those binary- and multi-class results expose that the NOR-class performance has been improved by $8.27\,\%$ margin with other constant experimental settings.
The identification results for the severity prediction (MoNCP vs. SeNCP) confer tremendous success in our pipeline, where barely $5.60\,\%$-MoNCP and $2.13\,\%$-SeNCP samples are prognosticated as SeNCP- and MoNCP-classes, respectively (see in Table~\ref{tab:confusion_folds}). 
Although overall macro-average AUC of the binary classification defeats the multi-class recognition (see in Fig.~\ref{fig:ROC_all_expts_folds}) by a margin of $2.1\,\%$, the later protocol has better class-balance results.
The multi-class protocol also provides less inter-fold variation than the binary-class, as depicted in Fig.~\ref{fig:ROC_all_expts_folds}. 
However, our approach for the COVID-19 identification exhibits praiseworthy achievement with high AUC values with less inter-fold variation in both of the class protocols.

\section{Conclusion}
\label{Conclusion}
During the current COVID-19 pandemic emergency, to mitigate the permanent lung damage due to coronavirus, precise recognition with negligible false negative is highly essential. This article aimed to design an artificial screening system for automated COVID-19 identification. A progressively resized 3D-CNN classifier is recommended in this study, incorporating lung segmentation, image augmentations, and class-rebalancing. The experimental analysis confirms that the CNN classifier's training with the suitable smaller patches and progressively increasing the network size enhance the identification results. Furthermore, incorporating the lung segmentation empowers the classifier to learn salient and characteristic COVID-19 features than utilizing whole chest CT images, driving to improved COVID-19 classification performance.
Again, the augmentations and class-rebalancing result in improved COVID-19 identification with high class-balanced recognition, shielding the network from being biased to a particular overrepresented class. 
In the future, the proposed pipeline will be employed in other volumetric medical imaging domain to validate its efficacy, versatility, and robustness. We also aim to deploy our trained model to a user-friendly web application for clinical utilization. The proposed system can be an excellent tool for clinicians to fight this deadly epidemic by the quicker and automated screening of the COVID-19.

\section*{CRediT authorship contribution statement}
\textbf{M. K. Hasan:} Conceptualization, Methodology, Software, Formal analysis, Investigation, Visualization, Writing- Review \& Editing, Supervision; 
\textbf{M. T. Jawad:} Software, Validation, Data Curation, Writing- Original Draft;
\textbf{K. N. I. Hasan:} Data Curation, Writing- Original Draft;
\textbf{S. B. Partha:} Writing- Original Draft;
\textbf{M. M. A. Masba:} Writing- Original Draft;

\section*{Acknowledgements}
None. No funding to declare. 

\section*{Conflict of Interest}
All authors have no conflict of interest to publish this research.

\bibliographystyle{model2-names}

\bibliography{sample}

\end{document}